\documentstyle[aps,twocolumn,epsfig]{revtex}

\begin{document}

\newcommand{\de}{^{o}}
\newcommand{\eq}{equation}
\newcommand{\beq}{\begin{equation}}
\newcommand{\eeq}{\end{equation}}
\newcommand{\bea}{\begin{eqnarray}}
\newcommand{\eea}{\end{eqnarray}}
\newcommand{\pri}{^{\prime}}
\def\alt{\,\raise 0.6ex\hbox{$<$}\kern -0.75em\lower 0.47ex
   \hbox{$\sim$}\,}
\def\agt{\,\raise 0.6ex\hbox{$>$}\kern -0.75em\lower 0.47ex
    \hbox{$\sim$}\,}
\newcommand{\real}{{\cal R}}
\newcommand{\bs}{{\vec{S}}}
\newcommand{\bls}{{\bf{s}}}
\newcommand{\bst}{{\bf{S}}_{tot}}

\newcommand{\dj}{\delta J}
\def\jour#1#2#3#4{{#1} {\bf #2}, #3 (#4).}
\def\tit#1#2#3#4#5{{#1} {\bf #2}, #3 (#4).}
\def\epl{Europhys. Lett.}
\def\prl{Phys. Rev. Lett.}
\def\pr{Phys. Rev.}
\def\prb{Phys. Rev. B}
\def\jpco{J. Phys. Cond. Mat}
\def\jpc{J. Phys. C}
\def\jap{J. Appl. Phys.}
\def\zpb{Z. Phys. B}



\twocolumn[\hsize\textwidth\columnwidth\hsize\csname    
@twocolumnfalse\endcsname                               

\begin{title} {\Large \bf
Frustrated Order by Disorder: the Pyrochlore Antiferromagnet
with Bond Disorder}

\end{title}

\author{L. Bellier-Castella$^{1}$, M.J.P. Gingras$^{2,3}$, P.C.W.
Holdsworth$^{4}$ and R. Moessner$^{5}$}
\address{$^1$Departement de Physique des Mat\'eriaux,
Universit\'e Claude Bernard, 69622 Villeurbanne, Cedex, France}
\address{$^2$Canadian Institute for Advanced Research}
\address{$^3$Department of Physics, University of Waterloo,
200 University Avenue West,
Waterloo,
Ontario, Canada N2L 3G1}
\address{$^4$Laboratoire de Physique, Ecole Normale Sup\'erieure,
46 All\'ee d'Italie, F-69364, Lyon, cedex 07, France}
\address{$^5$Department of Physics, Princeton University, Princeton, NJ 08544, USA
}

\vspace{3mm}

\date{\today}
\maketitle

\begin{abstract}

The classical Heisenberg antiferromagnet on the pyrochlore lattice is
macroscopically and continuously degenerate and the system remains
disordered at all temperatures, even in the presence of weak dilution
with nonmagnetic ions.  We show that, in stark contrast, weak {\em
bond} disorder lifts the ground state degeneracy in favour of locally
collinear spin configurations.  We present a proof that for a single
tetrahedron the ground state is perfectly collinear but identify two
mechanisms which preclude the establishment of a globally collinear
state; one due to frustration and the other due to higher-order
effects.  We thus obtain a rugged energy landscape, which is necessary
to account for the glassy phenomena found in real systems such as the
pyrochlore Y$_2$Mo$_2$O$_7$
recently reported by Booth {\em et al.}~\cite{Booth}  to
contain a substantial degree of bond disorder.

\end{abstract}

\vskip2pc]                                              





Geometrically frustrated magnets\cite{reviews},
on the kagom\'e or pyrochlore lattices, 
are ideal model systems for the study 
of general concepts pertaining to noncrystalline
condensed phases of matter. Behaviour analogous to that of 
liquid~\cite{Harris1}, glass~\cite{reviews,Y-exp,gaurei},
and ice~\cite{Harris2,Ramirez1} phases have already been attributed to
experimental pyrochlore systems, while the related
Gadolinium Garnet antiferromagnet offers liquid~\cite{GGG1} and re-entrant
solid behaviour~\cite{GGG2}. Equally, 
theoretical models provide the opportunity to study an array of many body 
phenomena:
the classical Heisenberg antiferromagnet on the pyrochlore lattice 
retains the behaviour of an idealized
liquid right to zero temperature~\cite{MC}, while in 
the equivalent kagom\'e~\cite{kagome} system
and in the XY antiferromagnet on the 
pyrochlore lattice~\cite{MC,Gingras1}  thermal ``order by disorder''  effects
lead to the apparition of liquid crystal like phases.

Given this rich array of phenomena one would like to present geometrically 
frustrated magnetism
as a field in which one can study charicatures of complex many body systems. 
However, 
despite some successes (see for example~\cite{reviews,GGG2,Gingras2}) 
there remain a number of marked disagreements
between experiment and theory which prevent such a desirable coherent picture. 
For example,
experiments on pure Y$_2$Mo$_2$O$_7$, which is believed to be well
represented by a Heisenberg antiferromagnet
on the pyrochlore lattice, show an extremely well defined spin-glass
transition~\cite{Y-exp}. Similar spin glass freezing is found in
Tb$_2$Mo$_2$O$_7$~\cite{gaurei}. This is in
sharp contrast to the behaviour of the classical 
Heisenberg antiferromagnet on the pyrochlore lattice, which 
shows no sign of glassy behaviour
at any temperature~\cite{MC}. 
In fact Y$_2$Mo$_2$O$_7$ behaves much more like a conventional
randomly disordered spin glass or even 
the disordered
compound CsNiCrF$_6$ -- which has two kinds of
magnetic ion (Cr$^{3+}$ and Ni$^{2+}$) arranged randomly
on a pyrochlore structure --
than that anticipated from a collective paramagnet.

Recent x-ray absorption measurements on Y$_2$Mo$_2$O$_7$~\cite{Booth} 
show that despite the high degree of stoichiometry
in this material, there is a small but measurable amount of disorder in 
the Mo$-$Mo bond lengths frozen into
the system at low temperature.  
In addition we note that the liquid like diffuse neutron scattering data 
observed for 
CsNiCrF$_6$ seem to be best
described by locally collinear spin configurations~\cite{Harris1}. 
Thermal selection of a collinear state has been considered
as a possibility in the Heisenberg pyrochlore antiferromagnet.
However, it has been shown 
not occur in the disorder-free~\cite{MC} and 
site-diluted cases~\cite{MB}.

In this paper we study the effect of weak bond disorder on the
classical Heisenberg antiferromagnet on the pyrochlore lattice. We
find that bond disorder provides a mechanism for the lifting of the the
ground state degeneracy, with the selection of a locally collinear
state. This kind of effect is known as configurational order by
disorder~\cite{VillainOBD}, since the ground states selected by the
perturbation typically incorporate long-range order.  However, we
demonstrate that configurational order by disorder is in turn
frustrated on this lattice, since the optimal configuration of
individual tetrahedra are not globally compatible, which leads to the
collinear order being only of short range. In addition, for finite
disorder strengths, the ground states even of individual tetrahedra
are no longer necessarily perfectly collinear. A rugged energy
landscape arises from the many ways of finding compromises between the
competing requirements imposed by the disorder. Our Monte Carlo
simulations do indeed find that the development of
local collinearity is accompanied by the onset of glassy behaviour.

The Heisenberg Hamiltonian for the pure pyrochlore antiferromagnet is
\begin{equation}
H_0 = J \sum_{<ij>} \vec S_i\cdot\vec S_j = {J\over{2}} \sum_p 
\vert \vec L_p \vert^2 - 
{NJ},
\end{equation}
where $J>0$ is the coupling constant, $N$ the number of spins and $\vec 
L_p$ is the
total spin on the $p^{th}$ tetrahedron. 
The sum on $<ij>$\ runs over all nearest neighbor bonds.
The ground state condition $\vec L_p\equiv 0$
leaves one degree of freedom per tetrahedron unconstrained,
thus yielding a
continuously degenerate ground state of extensive dimension $N/2$\ and
localised zero-energy degrees of freedom which fluctuate all the way
down to $T=0$~\cite{MC}.

Thermal order by disorder, in which ordered states are selected at
small but nonzero temperature due to the large entropy of the soft
fluctuations around them, has already been discussed for this 
model~\cite{MC,Reimers}.
The candidate states for selection
are the discrete subset of collinear ground states, which have the 
broken uniaxial symmetry of a nematic and are exponentially numerous
in $N$. However, localising the system in phase space near
points of this discrete set of collinear states is prohibitively
costly in entropy, even taking into account the soft fluctuations
around the collinear states: thermal order by disorder, unusually for
frustrated magnets, is absent here.

We add a small disordered term, $H_1$,
to $H_0$:
\begin{equation}
H_1= J\sum_{<i,j>} \delta_{i,j} \vec S_i \cdot \vec S_j\ ; \;\;\; 
-\delta_0 \;<\; \delta_{i,j} \;<\; \delta_0,
\end{equation}
with $\delta_{i,j}$ taken from a continous and rectangular distribution. 
We show, 
in the inset of Fig.~1,
the energy per spin at low temperatures (T/J =0.0006) for
different disorder strengths. As $\delta_0$ increases, the energy per
spin decreases linearly, illustrating that the disorder relieves the
frustration, allowing lower energy states for specific configurations.
One can see that these should be collinear by recognizing that the
energy on a given tetrahedron is minimized if the spins on the
weakest bonds (i.e those with $\delta_{i,j}$ most negative) are
ferromagnetically aligned, while those on the strongest bonds are
antiferromagnetically aligned. The energy of a single tetrahedron can
therefore be reduced compared to a non-collinear state by judicious
selection of which spins are up and which are down.  In the appendix
we give a proof that the lowest energy state is indeed a collinear
state for a single tetrahedron in the limit of weak disorder.

As a test for the formation of collinear order we have calculated 
the
probability density function $P(w)$ for the scalar product between
neighbouring spin vectors, $w_{i,j} = \vec S_i \cdot \vec S_j$,
by Monte
Carlo simulation.  For collinear states with 4 antiferromagnetic and 2
ferromagnetic bonds per tetrahedron, $P(w)$ is bimodal with delta
functions of strength $2/3$ and $1/3$ at $w= -1$ and $w = 1$
respectively.  In Fig.~1 we compare $P(w)$ for a system of 1024 spins
in a cubic cell with periodic boundaries for a system without disorder
and for a system with $\delta_0/J = 0.1$. In both cases, the data were
collected over $10^6$ Monte Carlo steps per spin (MCS) at $T/J =
0.001$ and the initial configuration generated by cooling in stages of
$10^5$\ MCS.  In the pure case on reducing the temperature a peak
develops at $w=-1$\ indicating the presence of weak local antiparallel
correlations.  In the presence of bond disorder the low temperature
feature at $w=-1$ is much more pronounced. In addition a clear peak in
probility is present also at $w=1$, as expected from the
configurational order by disorder. Nonetheless, collinear ordering
clearly remains incomplete.
\begin{figure}[h]
\vspace{-0.5cm}
\centerline{\epsfig{file=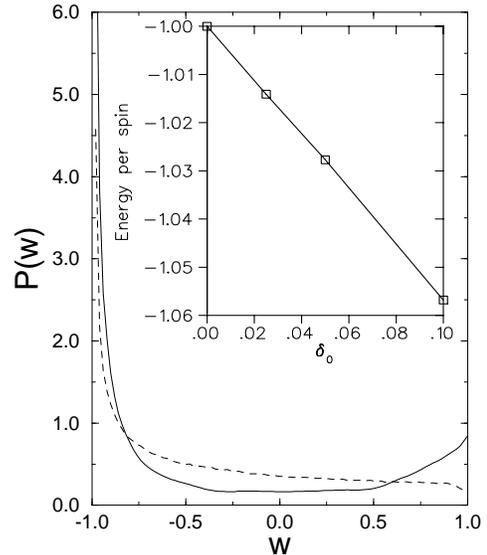,height=9cm}}
\vspace{-1cm}
\caption{Probability density function $P(w)$ for $1024$ spins. The
full line is for $\delta_0 = 0.1$ and the dotted line for
$\delta_0 = 0$.
The inset shows the
energy per spin with bond energy taken from the distribution (Eq.~2).
Values are from $N=1024$ spin systems averaged over 10 disorder 
configurations.}
\end{figure}
To see why this is the case, note that global collinear order by
disorder is frustrated by the presence of closed loops of bonds on the
lattice. If the single-tetrahedron ground states imply an odd number
of antiparallel bonds around any closed loop, a global ground state
obviously cannot be constructed consistently out of those for 
single tetrahedra.

Collinear order could be retained at the cost of having higher-energy
configurations on some of the tetrahedra, but this strategy is not
promising due to the large number of closed loops of bonds. The
alternative is to introduce canting, and this is indeed what happens,
as borne out by our simulations results.
The large number of possible compromises
imposed by the loops leads to a large number of states
with similar energy but separated by barriers - a rugged energy
landscape.

To leading order in $\delta_0$ the local energy minima lie on the
original ground state manifold with $\vec L_{p}=0$, while to higher
order they move off this surface towards the minima of $H_1$.  For
collinear states the spectrum of excitations of $H_0$ perpendicular to
this surface contains soft modes, and therefore canting away from
collinearity due to higher order effects can be considerable, even for
modest values of $\delta_0$. The higher order effects therefore provide
a mechanism for the destruction of collinearity, even for a single
tetrahedron and even for modest values of $\delta_0$.

We have tested for both spin glass behaviour and global collinearity,
since 
the ground state
manifold is now disconnected due to the formation of energy barriers
which could induce glassy behaviour and loss of ergodicity. 

The nematic
order parameter $Q$,  is the largest eigenvalue of the traceless
matrix~\cite{Birger}
\beq
 Q_{\alpha,\beta} = {3\over{2N}}\sum_i
 <(\vec S_i)_{\alpha}(\vec S_i)_{\beta}>  
-{1\over{2}}\delta_{\alpha\beta},
\eeq
where $\alpha = x,y,z$ and $\delta_{\alpha\beta}$\ is the 
Kronecker delta. The Edwards-Anderson spin 
glass order parameter~\cite{FH} is
\beq
q_{EA} = {1\over{N}}\sum_i <\vec S_i>^2.
\eeq
\begin{figure}[h]
\vspace{0cm}
\centerline{\epsfig{file=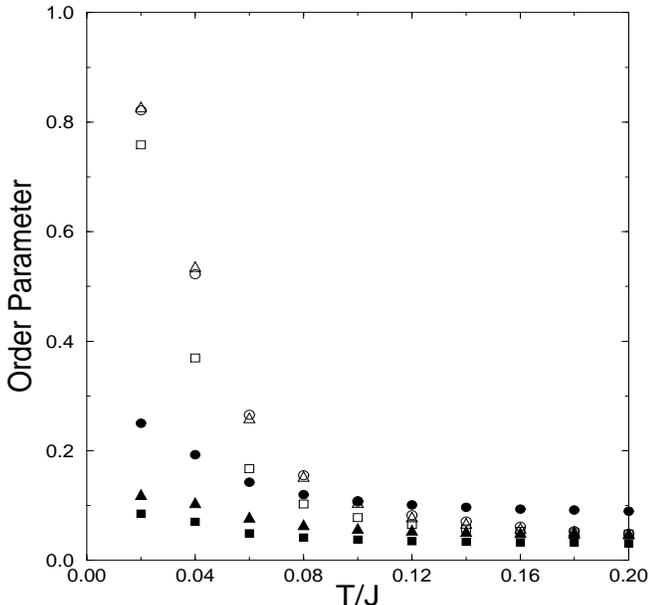,width=7cm,height=7cm}}
\vspace{0.5cm}
\caption{{ Nematic order parameter $Q$ (filled symbols)and 
spin glass order parameters $q^{EA}$ (open symbols) against
temperature, calculated for 
128 spins (circles), 432 spins (triangles) and 1024 spins (squares). }}
\end{figure}

In Fig.~2 we show $Q$ and $q_{EA}$ against temperature for
$\delta_0=0.1$ for $N=128,432$ and $1024$ spins. The data is averaged
over $10-30$ configurations of disorder, with at each temperature
$10^5$ MCS for equilibration and $10^5$ for production. The data
clearly show the onset of glassy behaviour at a temperature of the
order of $T/J= \delta_0/J=0.1$.  There is a marked decrease of
$q_{EA}(T)$ for $T < 0.08$ from $N=432$ to $N=1024$, suggesting that
the freezing occurs over microscopic clusters only. Moreover,
preliminary results for the susceptibility suggest, at least for small
systems, that there is no bifurcation between field-cooled and zero
field cooled measurements. However, the time scales involved do
increase by orders of magnitude in the presence of bond disorder: the
relaxation time at the lowest temperature studied, $T/J=0.02$ is in
excess of $10^6$ MCS, while for the pure system it is of order $10^3$
MCS.
 
The nematic order parameter scales to zero at
low temperature roughly as $Q(T=0) \sim 4/\sqrt{N}$ 
confirming that the collinearity is only a local phenomena, with
correlations over about 4 tetrahedra.

We conclude that, although the ground state for a single tetrahedron
is perfectly ordered, the state attained for the larger system is only
partially ordered and is dynamically frozen on the time scales we have
considered.  This may not only be because the true ground state is not
perfectly collinear, as discussed above, but also because a loss of
ergodicity in the simulations can mean that the system gets stuck in a
metastable state.

Our results therefore provide a plausible mechanism for the local
collinearity proposed to fit the neutron scattering data of Harris et
al on CsNiCrF$_6$~\cite{Harris1}. Bond disorder alone with isotropic
Heisenberg spins 
is probably not sufficient to explain the
glassy behaviour observed in Y$_2$Mo$_2$O$_7$. However it is sufficient
to remove the anomalous liquid like behaviour
of the classical Heisenberg antiferromagnet
and ensure dynamical freezing at a well defined temperature, for any 
finite system. One might naively expect that, due to absence of
{\it long-range} nematic order, the present system renormalizes
to an effective randomly frustrated 
isotropic Heisenberg spin glass, at length scales
beyond the short-range correlated collinear regions.
It is generally believed
that the isotropic three dimensional Heisenberg spin glass has a 
zero temperature
spin-glass freezing transition~\cite{olive}, 
with no field-cool/zero field-cool split in
the magnetization as is also indicated by our simulations.
Within this picture,
the  observed spin glass transition in
Y$_2$Mo$_2$O$_7$~\cite{Y-exp}		
must 
come from anisotropic perturbations~\cite{gingras-prl}.
Site disorder has also been considered
previously~\cite{MB}, but it does not lead to glassy behaviour:
substitution of a magnetic ion does not destroy the continuous
ground state manifold but only reduces its dimension.
 Bond disorder, on the other
hand, should lead to energy barriers of a height set by
the disorder
strength, which is just what we find in Fig.~2.
More work is clearly
required here. In particular,  further simulations are needed to investigate the
nature of the glass transition in more
detail. It would also be of interest to re-examine neutron data with a
view to fitting $S(q,\omega)$ using local collinear spin configurations. 

Finally we note that the longest standing theoretical challenge in
this field is motivated by the most studied experimental system
SrCr$_x$Ga$_{12-x}$O$_{19}$ (SCGO) which consists of bi-layers of two
kagom\'e lattices of magnetic ions which are coupled, forming a slice
of pyrochlore lattice.  There is experimental evidence that, at low
temperature the spins form into a glassy co-planar
phase~\cite{reviews,SCGO,schiffer} but no theoretical explanation for
this result exists.  The experimental compound is non-stoichiometric
and so substitutional disorder is inevitable. The thermal order by
disorder seen in the classical kagom\'e~\cite{kagome} antiferromagnet
has been shown to be extremely sensitive to disorder~\cite{SCHB}, with
2\% dilution being sufficient to destroy coplanarity, and so cannot be
responsible for the observed coplanarity. Further to this, in presence
of the $12k-2a-12k$ kagom\'e$-$triangular$-$kagom\'e coupling, the
coplanar thermal order-by-disorder found for a single kagom\'e
layer is destroyed, with or without site or bond
disorder.  It is therefore tempting to speculate that a variant of the
mechanism presented here could lead to configurational coplanar order
by disorder in parallel with glassy behaviour, in agreement with
experiment at long last. However, we have not, as yet, managed to
isolate such a variant.

\newpage

\acknowledgements

\noindent This reserach has been funded by
NSERC of Canada and the Pole Scientifique de Mod\'elisation 
Num\'erique (PSMN) of the Ecole Normale Superieure de Lyon.
M.G. and P.H. are grateful to the Association of Universities and Colleges of
Canada for a travel grant from the Going Global STEP program.
M.G. acknowledges
the Research Corporation for a Research Innovation Award and a Cottrell
Scholar Award, and the Province of Ontario for a Premier Research
Excellence Award.

\appendix
\setcounter{equation}{0}
\section{}\label{app.1}

We consider the Hamiltonian
$H=H_0+H_1$ for a single tetrahedron, and treat $H_1$
in first order of perturbation theory.
The ground states of $H_0$\ have total spin $\vec L=0$. They can be
parametrized 
as 
$\bs_1=(0,0,1); \
\bs_2=(0,\sin \alpha, \cos \alpha);\
\bs_3=(\sin \beta \sin \phi,\sin \beta \cos \phi, \cos \beta);\ 
\bs_4=-\bs_1-\bs_2-\bs_3 .
$
The normalisation of $\bs_4$\ imposes the restriction
$
\cos\alpha+\cos\beta\leq0
\label{eq:restr},
$
as well as a condition on $\phi$~\cite{reimers}.
Therefore, 
\bea
\bs_1\cdot\bs_2 & =& \cos\alpha	\;\;,	\nonumber \\
\ \bs_1\cdot\bs_3 & = &\cos\beta \;\;,	\nonumber \\
\bs_1\cdot\bs_4 & = &-1-\cos\alpha-\cos\beta \;\; .
\nonumber
\eea

In a ground state of $H_0$, the sum of any pair of spins 
is opposite to the sum of the other pair of spins, and hence 
\bea
\bs_3\cdot\bs_4=\bs_1\cdot\bs_2\;\;,\;\;
\bs_2\cdot\bs_4=\bs_1\cdot\bs_3\;\;,\;\;
\bs_2\cdot\bs_3=\bs_1\cdot\bs_4 \;\;.\nonumber
\eea
The dot products between two spins can
thus
be expressed using
$\cos\alpha$\ and $\cos\beta$\ only, this means that, restricted to
the ground states of $H_0$, $H_1$\ must have the following form: 
\bea
H_1/J=A \cos\alpha + B\cos\beta+C \;\;, 
\eea 
where $A, B$\ and $C$\ depend on
the $\delta$: $A=\delta_{1,2}+\delta_{3,4}-\delta_{1,4}-\delta_{2,3}$,
$B=\delta_{1,3}+\delta_{2,4}-\delta_{1,4}-\delta_{2,3}$, and 
$C=-\delta_{1,4}-\delta_{2,3}$.

We now have to distinguish the following cases:
(I) $A\neq B$. Without loss of generality, let $A>B$. (a) For
$A>0,B\neq0$, the global minimum of $H_1$\ is given by $\cos\alpha=-1$\ and
$\cos\beta=-{\rm sign}(B)$. (b) If $A>0,B=0$, then $\cos\beta$\ is
undetermined. (c) For $A<0$, $H_1$\ is minimized by $\cos\alpha=-1,
\cos \beta=1$.
(II) $A=B$ (a) $A>0$: $H_1$\ is minimized by $\cos\alpha=\cos\beta=-1$.  
(b) $A<0$: $H_1$\ is minimized by $\cos\alpha+\cos\beta=0$.
(c) $A=0$: $H_1$\ is independent of $\alpha, \beta$. 

From this, we see that two things can happen. Either all spins are
collinear, in the cases Ia,Ic,IIa, or the degeneracy is not fully
lifted, in the special (finetuned) cases Ib,IIb,IIc. For the latter, a
collinear state still exists which minimizes the energy. In other
words, a non-collinear state is never selected by the disorder as
unique ground state, and the generic case $A\neq B, AB\neq0$\ has a
unique ground state which is collinear.



\newpage


\begin{thebibliography}{99}

\bibitem{Booth}C.H. Booth, J.S. Gardner, G.H. Kwei, R.H. Heffner,
F. Bridges and M.A. Subramanian, to appear in Phys. Rev. B. 

\bibitem{reviews}
For reviews, see: A. P. Ramirez, Annu. Rev. Mater. Sci. {\bf 24},
453, (1994); 
P. Schiffer and A. P. Ramirez, \tit{Comments
Cond. Mat. Phys.}{18}{21}{1996}.

\bibitem{Harris1} M.J. Harris, M.P. Zinkin, Z.Tun, B.M. Wanklyn and
I.P.  Swainson, 
Phys. Rev. Lett. {\bf 73}, 189 (1994);
M.P. Zinkin, M.J. Harris
and T. Zeiske, \tit{\prb}{56}{11786}{1997}.

\bibitem{Y-exp} S.R. Dunsiger et al.,
Phys. Rev. B {\bf 54}, 9019 (1996);
M.J.P. Gingras et al.,
Phys. Rev. Lett. {\bf 78}, 947 (1997);
J.S. Gardner et al.,
Phys. Rev. Lett. {\bf 83}, 211 (1999).

\bibitem{gaurei} B. D. Gaulin, J. N. Reimers, T. E. Mason, J. E. Greedan and Z.
Tun,
\tit{\prl}{69}{3244}{1992}.

\bibitem{Harris2} M.J. Harris, S.T. Bramwell, D.F. McMorrow, T. Zeiske
and K.W.  Godfrey, \tit{\prl}{79}{2554}{1997}.

\bibitem{Ramirez1} A. Ramirez, A. Ayashi, R. Cava, R. Siddarthan and
B. Shastry, Nature {\bf 399}, 333 (1999).

\bibitem{GGG1} O.A. Petrenko, C. Ritter, M. Yethiraj and D. McK Paul,
\tit{\prl}{80}{4570}{1998}.

\bibitem{GGG2} P. Shiffer, A.P. Ramirez, D.A. Huse and A.J. Valentino, 
\tit{\prl}{73}{2500}{1994}.

\bibitem{MC} R. Moessner and J.T. Chalker, 
Phys. Rev. Lett. {\bf 80}, 2929 (1998);
\tit{\prb}{58}{12049}{1998}.

\bibitem{kagome} J.T. Chalker, P.C.W. Holdsworth and E.F. Shender, 
Phys. Rev. Lett. {\bf 68}, 855 (1992);
J.N. Reimers and A.J. Berlinsky, 
Phys. Rev. B {\bf 48}, 9539 (1993); I. Ritchey, P. Chandra and 
P. Coleman, Phys. Rev. B {\bf 47}, 15342 (1993);
A.B. Harris, C. Kallin 
and A.J. Berlinksy, \tit{\prb}{45}{2899}{1992}.

\bibitem{Gingras1} S.T. Bramwell, M.J.P. Gingras and J.N. Reimers, 
J. Appl. Phys. {\bf 75}, 5523 (1994).


\bibitem{Gingras2} B. C. den Hertog and M. J. P. Gingras,
\tit{\prl}{84}{3430}{2000}.


\bibitem{MB} R. Moessner and A.J. Berlinsky, \tit{\prl}{83}{3293}{1999}.

\bibitem{VillainOBD} J. Villain, R. Bidaux, J.P. Carton and R.J. Cont\'e, 
J. Phys. (Paris) {\bf 41}, 1263 (1980).

\bibitem{Reimers}J.N. Reimers, \tit{\prb}{45}{7287}{1992}.

\bibitem{reimers}J. N. Reimers, A. J. Berlinsky and A.-C. Shi,
\tit{\prb}{43}{865}{1991}.


\bibitem{Birger} M. Plischke and
B. Bergersen, ``Equilibrium Statistical Physics'', World Scientific, 1994.

\bibitem{FH} K.H. Fischer and J.A. Hertz, ``Spin Glasses'', Cambridge University
Press, Cambridge (1991).


\bibitem{olive} J.A. Olive, A.P. Young and D. Sherrington, 
Phys. Rev. B {\bf 34}, 6341 (1986). 

\bibitem{gingras-prl}
M.J.P. Gingras,
Phys.  Rev. Lett. {\bf 71}, 1637
(1993).

\bibitem{SCGO}
A. P. Ramirez, G.P. Espinosa and A.S. Cooper, \tit{\prl}{64}{2070}{1990}{***} 


\bibitem{schiffer}
P. Schiffer,
A. P. Ramirez, K. N. Franklin and S-W. Cheong,
Phys. Rev. Lett. {\bf 77}, 2085 (1996).

\bibitem{SCHB} E.F. Shender, V.B. Cherepenov, P.C.W. Holdsworth and
A.J. Berlinsky, \tit{\prl}{70}{3812}{1993}.




\end{thebibliography}
\end{document}